\documentclass[apj]{emulateapj}

\def\lesssim{\mathrel{\hbox{\rlap{\hbox{\lower4pt\hbox{$\sim$}}}\hbox{$<$}}}}
\def\gtrsim{\mathrel{\hbox{\rlap{\hbox{\lower4pt\hbox{$\sim$}}}\hbox{$>$}}}}

\def\gax{\mathrel{\raise.3ex\hbox{$>$}\mkern-14mu\lower0.6ex\hbox{$\sim$}}}
\def\lax{\mathrel{\raise.3ex\hbox{$<$}\mkern-14mu\lower0.6ex\hbox{$\sim$}}}
\def\gtorder{\mathrel{\raise.3ex\hbox{$>$}\mkern-14mu
             \lower0.6ex\hbox{$\sim$}}}
\def\ltorder{\mathrel{\raise.3ex\hbox{$<$}\mkern-14mu
             \lower0.6ex\hbox{$\sim$}}}

\begin{document}

\title{The All-Sky Automated Survey for Supernovae (ASAS-SN) Light Curve Server v1.0}

\author{C.~S. Kochanek\altaffilmark{1,2}, 
   B.~J.~Shappee\altaffilmark{3,4},
   K.~Z.~Stanek\altaffilmark{1,2},
   T.~W.-S. Holoien\altaffilmark{1,2,8},
   Todd~A. Thompson\altaffilmark{1,2},
   J.-L. Prieto\altaffilmark{5,6},
   Subo Dong\altaffilmark{7},
   J.~V.~Shields\altaffilmark{1},
   D.~Will\altaffilmark{1},
   C.~Britt\altaffilmark{9},
   D.~Perzanowski\altaffilmark{9},
   G.~Pojma\'nski\altaffilmark{10}
 }

\altaffiltext{1}{Department of Astronomy, The Ohio State University, 140 West 18$^{th}$ Avenue,
  Columbus, OH 43210} 

\altaffiltext{2}{Center for Cosmology and Astroparticle Physics, The Ohio State University,
  191 W. Woodruff Avenue, Columbus, OH 43210} 

\altaffiltext{3}{Carnegie Observatories, 813 Santa Barbara Street, 
   Pasadena, CA 91101} 

\altaffiltext{4}{Hubble and Carnegie-Princeton Fellow} 

\altaffiltext{5}{ N\'ucleo de Astronom\'ia de la Facultad de Ingenier\'ia y Ciencias, Universidad Diego Portales, Av. Ej\'ercito 441, Santiago, Chile}

\altaffiltext{6}{Millennium Institute of Astrophysics, Santiago, Chile }

\altaffiltext{7}{Kavli Institute for Astronomy and Astrophysics, Peking University, Yi He Yuan Road 5, Hai Dian District, China}

\altaffiltext{8}{US Department of Energy Computational Science Graduate Fellow}

\altaffiltext{9}{ASC Technology, College of Arts and Science, The Ohio State University, 125 S. Oval Mall, Columbus, OH 43235}

\altaffiltext{10}{Warsaw University Observatory, Al Ujazdowskie 4, 00-478Warsaw, Poland}

\begin{abstract}
The All-Sky Automated Survey for Supernovae (ASAS-SN) is working towards imaging the 
entire visible sky every night to a depth of $V\sim17$~mag. The present data covers the
sky and spans $\sim2$-$5$~years with $\sim100$-$400$ epochs of observation.
The data should contain some $\sim 1$ million variable sources, and the 
ultimate goal is to have a database of these observations publicly accessible.
We describe here a first step, a simple but unprecedented web interface 
{\tt https://asas-sn.osu.edu/}
that 
provides an up to date aperture photometry light curve for any user-selected sky 
coordinate.  Because the light curves are produced in real time, this web tool is 
relatively slow and can only be used for small samples of objects.  However, it also 
imposes no selection bias on the part of the ASAS-SN team, allowing the user to 
obtain a light curve for any point on the celestial sphere.  We present the tool, describe its 
capabilities, limitations, and known issues, and provide a few illustrative examples. 
\end{abstract}


\section {Introduction}

\cite{Paczynski2000} outlined the astrophysical need for all-sky variability
surveys and the desirability of public access. Well over a decade later, the 
All-Sky Automated Survey for Supernovae (ASAS-SN, \citealt{Shappee2014}) is the first to routinely 
survey the entire visible sky, reaching a depth of roughly $17$~mag.  ASAS-SN has publicly 
announced all transient detections from its inception.  Here we take the first 
step towards making the data more publicly accessible, with a 
tool to allow astronomers to obtain a current light curve of an arbitrary point 
on the sky.
  
Many ground-based surveys have offered retrospective access to databases of 
light curves of varying depths and survey areas.  The widest area examples
are the original All-Sky Automated Survey (ASAS, \citealt{Pojmanski2002}), 
the Northern Sky Variability Survey (NSVS, \citealt{Wozniak2004}),
the Catalina Real-Time Survey (CRTS, \citealt{Drake2009}) and the
Palomar Transient Factory (PTF, \citealt{Law2009}).  Examples of 
smaller area surveys with publicly available light curves 
are the Optical Gravitational Lensing Experiment
(OGLE, \citealt{Udalski2008}) and the Sloan Digital Sky Survey's
Stripe 82 (SDSS, \citealt{Ivezic2007}).  For many of the brightest
variables, the compilation of data from (primarily) amateur 
astronomers by the American Association of Variable Star Observers
(AAVSO) remains the best public source of data.

As of mid-2017, ASAS-SN consists of two stations, located at the 
Haleakala Observatory (Hawaii) and the Cerro Tololo International
Observatory (CTIO, Chile) sites.  The stations are hosted by 
Las Cumbres Observatory (\citealt{Brown2013}), and are operated
through their network.  By
the end of 2017, ASAS-SN will have five stations, with the addition of 
a second unit at CTIO and one each at McDonald Observatory (Texas), 
and the South African Astrophysical
Observatory (SAAO, Sutherland, South Africa) sites.  The new
stations will also be hosted by Las Cumbres Observatory.  

A station consists
of four 14~cm aperture Nikon telephoto lenses, each with a thermo-electrically cooled, 
back-illuminated, 2048$^2$, Finger Lakes Instruments, ProLine CCD camera.  The 
field of view of each camera is roughly 4.5~deg$^2$, the pixel scale is 
8\farcs0, and the image FWHM is $\sim 2$ pixels. With overlaps 
between observing fields, the instantaneous ``field of view'' with
5 stations will be roughly
360 square degrees.  Observations are made using the V (original two stations)
or g (three new stations) band filters and three dithered 90~sec exposures.
Assuming an average 10 hour night, the system will be able to survey 48,000 
square degrees per night by the end of 2017, with considerable robustness
against weather losses thanks to the multiplicity of sites. The operations
of ASAS-SN are presently funded through the end of 2021.

\begin{figure*}
\centering
\includegraphics[width=6.0in]{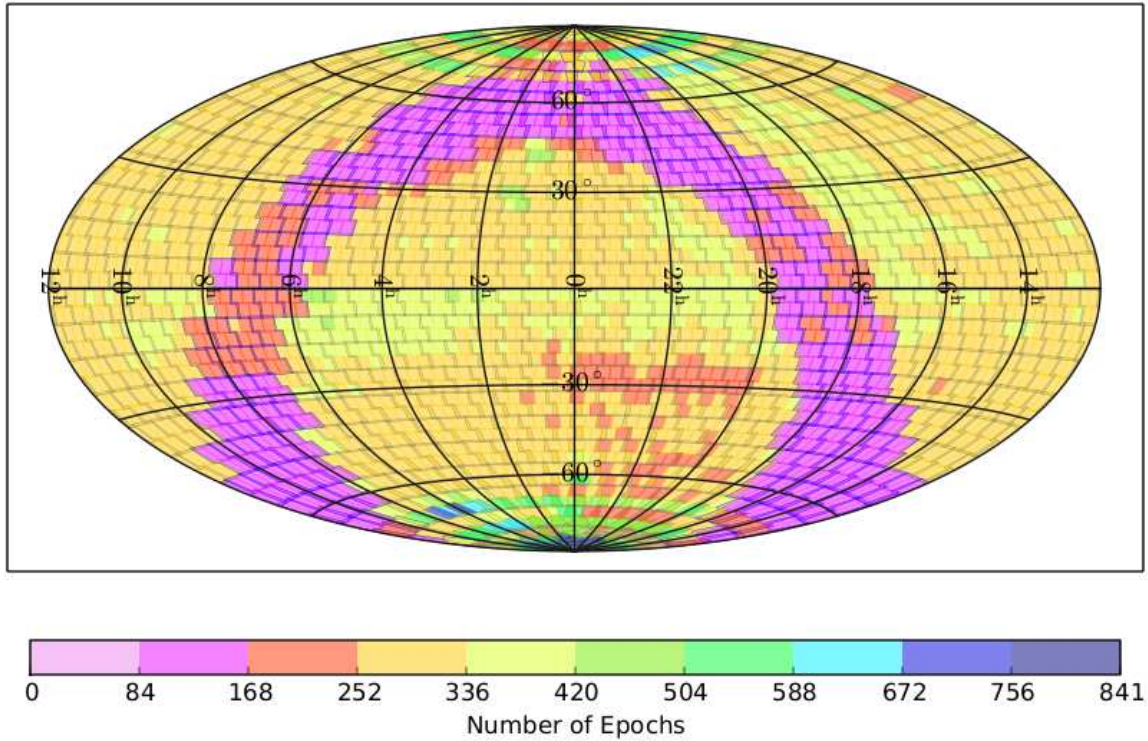}
\caption{ Equatorial projection of the number of available ASAS-SN epochs.  There are some artifacts
  because of historical details as ASAS-SN built up to two complete
  stations.  There are, on average, $2.6$ images for each epoch.
  }
\label{fig:epochs}
\end{figure*}

To date, ASAS-SN has focused on its primary, original goal of carrying out
a survey for bright transients across the visible sky with minimal observational bias. 
This was particularly aimed at obtaining a complete inventory of nearby supernovae (SNe)
to well characterize the local SN rate and its correlations with galaxy type
and properties.  The supernova search, cataloged in \cite{Holoien2017a,Holoien2017b,Holoien2017c},
has led to a significant increase in the discovery rate of bright SNe while 
eliminating the bias of amateurs towards
SNe in large galaxies. ASAS-SN is also finding many more SNe close to the 
cores of galaxies than other amateur and professional searches.  

In the process of finding SNe, including the most luminous SN to date
(ASASSN-15lh, \citealt{Dong2016}, \citealt{Godoy2017}),
many other transient sources have also been routinely discovered.
The most common are cataclysmic variables (CVs), which are so numerous that they
are simply released on a separate WWW page\footnote{
  http://www.astronomy.ohio-state.edu/$\sim$assassin/transients.html}
and the light curves are available from 
 the ASAS-SN CV Patrol\footnote{http://cv.asassn.astronomy.ohio-state.edu/} (\citealt{Davis2015}).
ASAS-SN is the dominant source for reporting new CVs and CV outbursts 
(see, e.g., \citealt{Kato2016}).  Rarer Galactic events are classical novae
(e.g., \citealt{Stanek2016}), M and even L dwarf flares
(e.g., \citealt{Schmidt2014,Schmidt2016}), and
outbursts of young stellar objects (\citealt{Holoien2014a}, \citealt{Herczeg2016}).  Rarer 
extragalactic events are tidal disruption events (TDEs), where the
majority of the best studied TDEs have been discovered by ASAS-SN
(\citealt{Holoien2014b,Holoien2016a,Holoien2016b}, 
\citealt{Brown2016,Brown2017a})
or are present in ASAS-SN (\citealt{Brown2017b}), and ``changing-look''
AGN (\citealt{Shappee2014}).

ASAS-SN has not focused on other sources of variability to date, 
although the project has supplied light curves or variability 
searches for a range of other projects.  For example, there 
are studies of ``eclipsing'' events 
(\citealt{Rodriguez2016}, \citealt{Rodriguez2017}, \citealt{Osborn2017}) 
and other (\citealt{Gully2017}) properties of T Tauri stars. 
\cite{Littlefield2016} examined the recovery of the intermediate polar
FQ~Aquarii from its low state in 2016.  ASAS-SN observations were
also used to limit any optical counterparts to high energy $\gamma$-ray 
(\citealt{Abseysekara2015}) and ICECUBE neutrino events (\citealt{Aartsen2017}).
Internally, ASAS-SN has identified large numbers of previously 
unreported variable stars which will start being released in the
near future.   

As part of the next phase of ASAS-SN, the intent is to
provide a steadily expanding set of tools for obtaining ASAS-SN light curves.
The first version, released here, is a tool to obtain an ASAS-SN
light curve of any user-selected position of the sky.  It is a relatively slow tool
since it simply carries out aperture photometry at the requested location,
but it also involves no preconceptions on the part of ASAS-SN
as to what represents an interesting source.
Bulk light curve requests will not be possible with the current tool because
the necessary computations are done on request -- we
invite anyone interested in large numbers of light curves to contact the
ASAS-SN team.  The next phase will be to build a database of the light curves of known variables
and variables discovered by ASAS-SN, with the goal of having this tool available
in 2018.  The final phase will be to have a complete database of light
curves for ASAS-SN sources.  The staged releases will allow the ASAS-SN
team to develop, test, and debug the releases in a logical and controlled 
order from the simple to the complex.  

In \S2 we describe the initial tool
and present some examples of both its uses and its limitations.  In \S3
we provide a short summary.  
Research making use of these light curve tools either through direct
use of the light curves or simply as a source for confirmations
should cite \cite{Shappee2014}
for the ASAS-SN survey and this paper for the light curves.

\begin{figure*}
\centering
\includegraphics[width=6.0in]{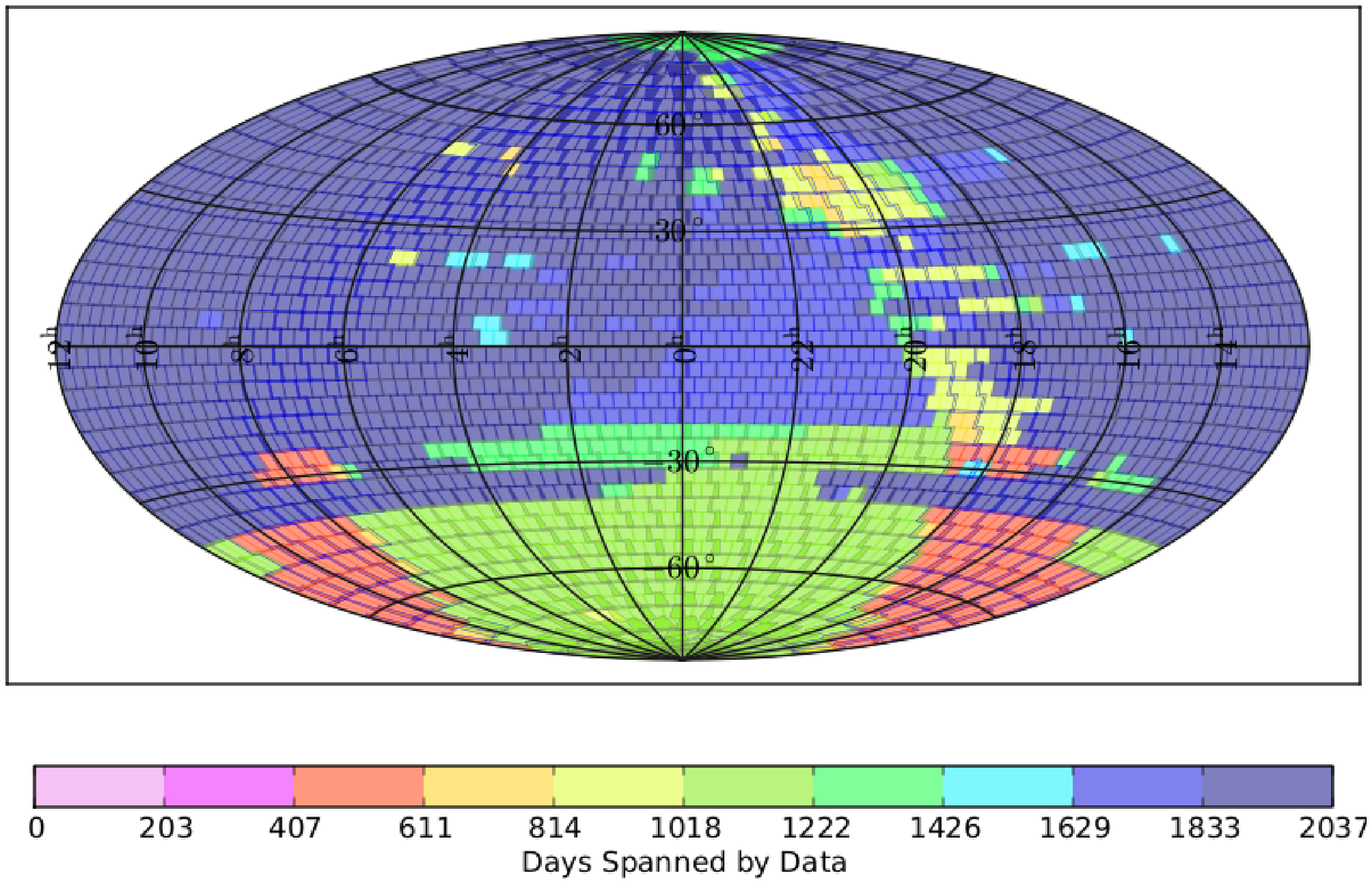}
\caption{ Equatorial projection of the time spanned by the ASAS-SN data.  There are some artifacts
  because of historical details as ASAS-SN built up to two complete stations,
  and the temporal coverage of the (Northern) Galactic plane has a significant gap.
  }
\label{fig:span}
\end{figure*}

\begin{figure*}
\centering
\includegraphics[height=2.8in]{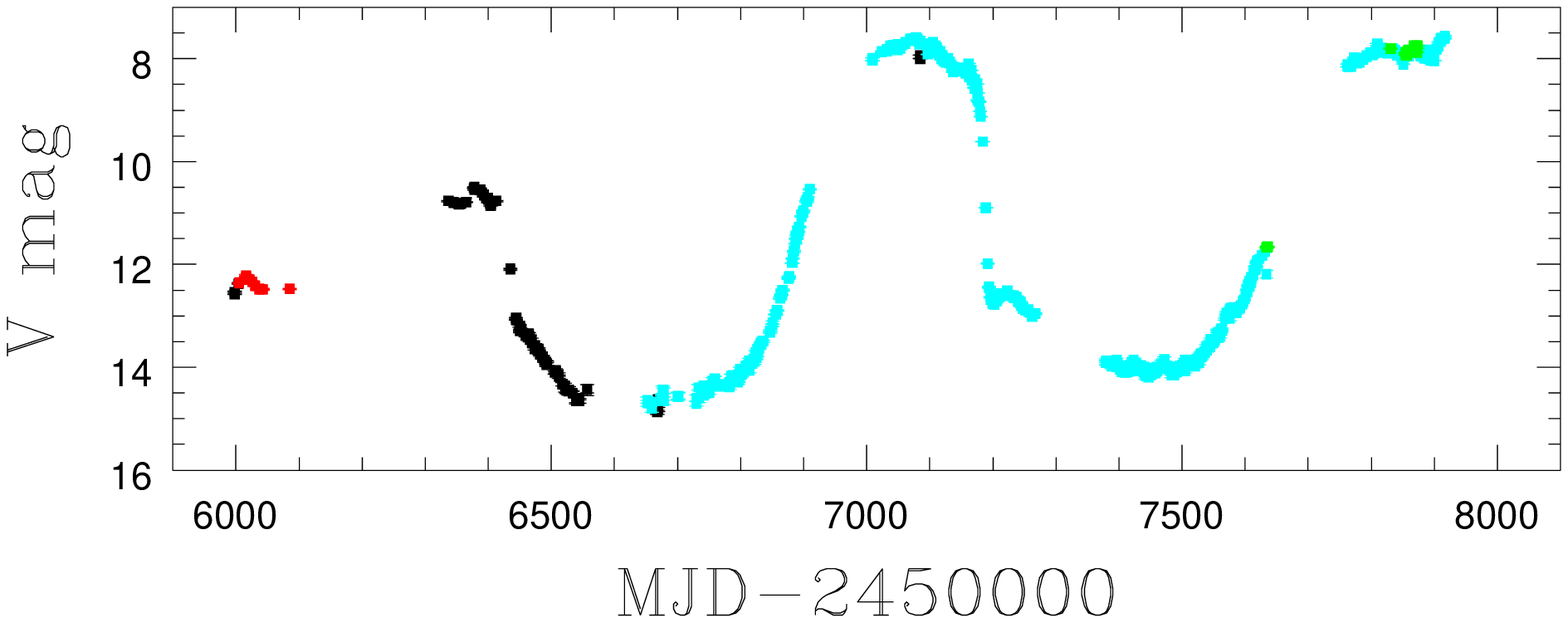}
\caption{ The ASAS-SN light curve of R Coronae Borealis.  The different colors
  are data from different cameras.
  }
\label{fig:rcorbor}
\centering
\includegraphics[height=2.8in]{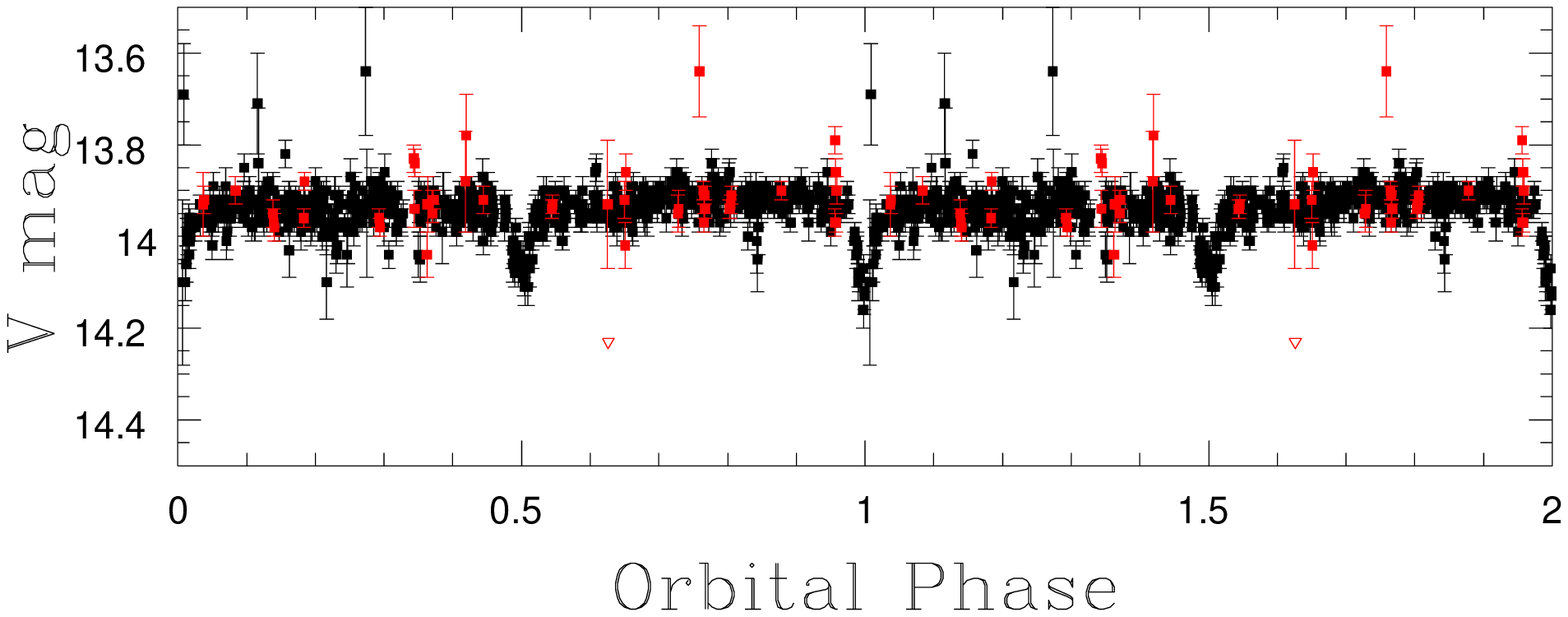}
\caption{ Phased ASAS-SN light curve of the eclipsing binary pair of M dwarfs
  KELT~J041621$-$620046 (\citealt{Lubin2017}). The different colors are data from 
  different cameras. 
  }
\label{fig:mdwarf}
\centering
\includegraphics[height=2.8in]{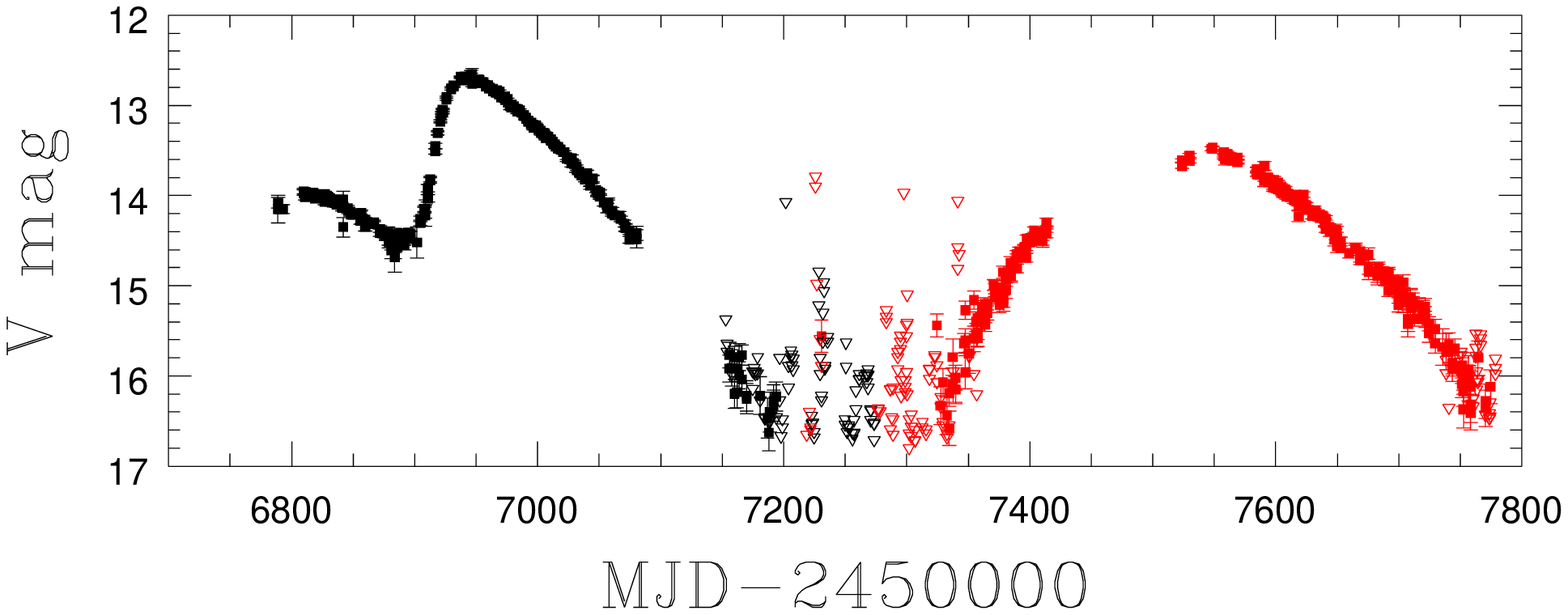}
\caption{ ASAS-SN light curve of the proposed Thorne-Zytkow object HV~2112 in
  the Small Magellanic Cloud (\protect\citealt{Levesque2014}).  Squares with error 
  bars are detections and triangles are $5\sigma$ upper limits.  
  The different colors are data from different cameras.  
  }
\label{fig:tzo}
\end{figure*}

\begin{figure*}
\centering
\includegraphics[height=2.8in]{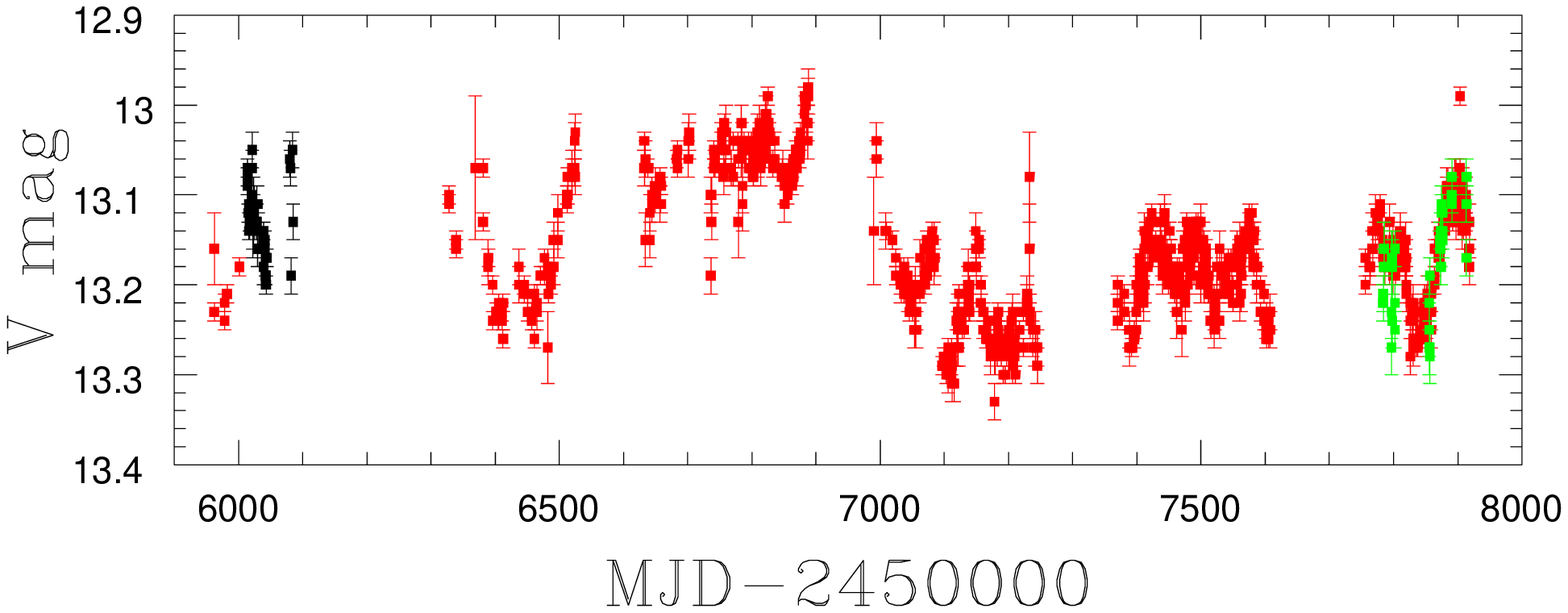}
\caption{ ASAS-SN light curve of the AGN NGC~5548. 
  One anomalous point falls below the bottom of the figure.
  The different colors
  are data from different cameras.
 }
\label{fig:ngc5548}
\vspace{0.10in}
\centering
\includegraphics[height=2.5in]{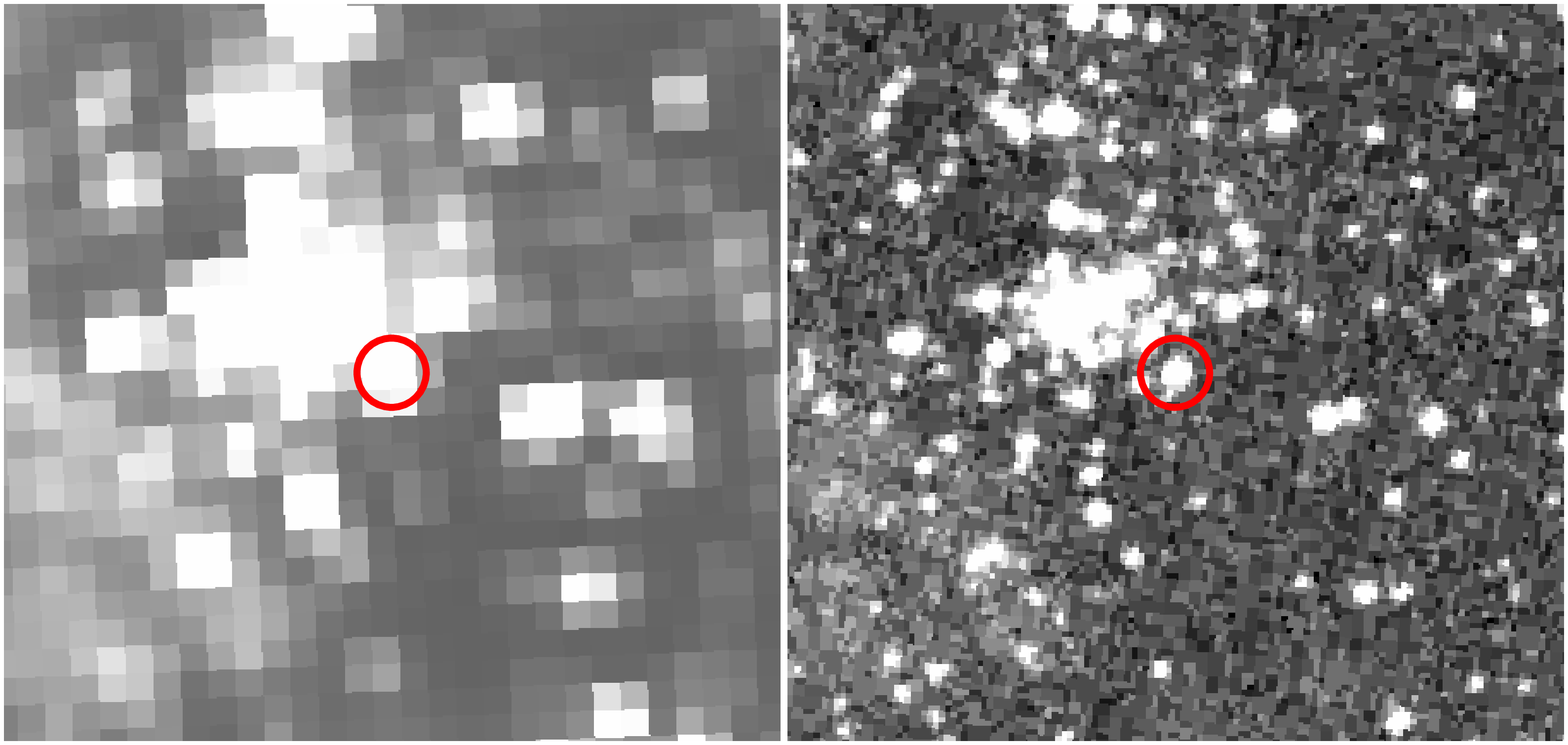}
\caption{ ASAS-SN (left) and DSS (right) images of the field of
  WR20a, where the circle around the source is $10\farcs0$ in radius. 
  Contamination due to the local crowding will generally distort an 
  ASAS-SN light curve by an additive constant. 
  }
\label{fig:crowd1}
\centering
\includegraphics[height=2.8in]{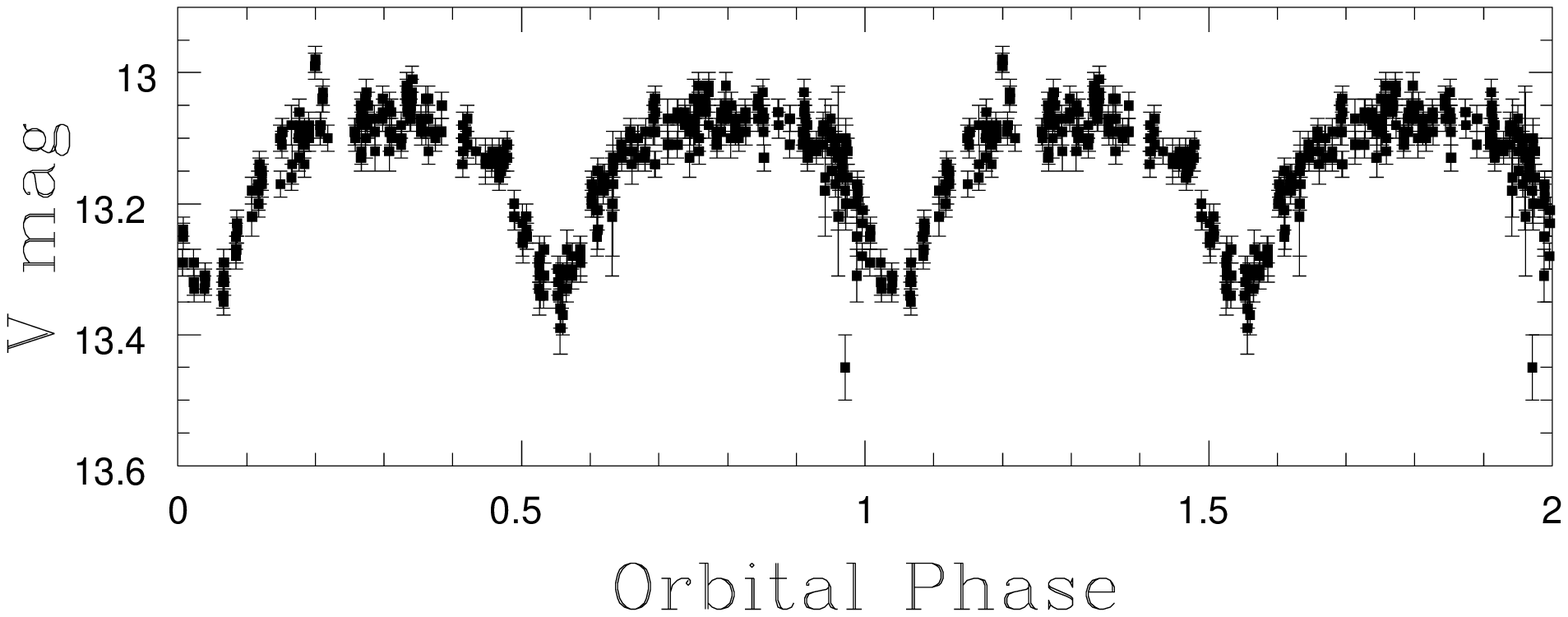}
\caption{ Phased ASAS-SN light curve of WR20a using the
  primary eclipse time of 2453124.569 from \cite{Bonanos2004}
  and a revised period of 3.684599~days. Blending has reduced
  the amplitude of the eclipses by about $0.1$~mag.  
  One bad point lies 
  off the bottom of the figure.
  }
\label{fig:crowd2}
\end{figure*}
\section{ASAS-SN Light Curves}

The requirement that we impose no restriction on the location at
which a light curve can be obtained means that the light curves 
must be computed at the time of the request.  This is a 
straightforward process, but slow compared to pre-computing light curves for
a defined sample of sources and storing them in a database. As
discussed in the introduction, such tools will be made available for
ASAS-SN in the relatively near future.  

Figures~\ref{fig:epochs} and \ref{fig:span} summarize the number
of epochs and the temporal span of the ASAS-SN data at the 
present time.  On average there are $2.6$ images per
epoch, which is less than the nominal value of $3$ because we 
initially used only $2$ dithered images per epoch and weather 
and scheduling idiosyncrasies can lead to obtaining fewer than
$3$ good images.
Most of the structures in the two Figures can be understood from
the history by which ASAS-SN reached its current state with two
stations each with four cameras.  ASAS-SN started with the
Hawaii station and two cameras ({\tt ba} and {\tt bb})
in early 2013, and it was expanded to four cameras (adding {\tt bc}
and {\tt bd}) in (roughly) December 2013.  The 
CTIO station was deployed with two cameras ({\tt be} and {\tt bf})
in May 2014 and expanded to four cameras (adding {\tt bg} and
{\tt bh}) in July 2015.  

Initially, we tried to observe the 
entire sky visible from Hawaii, but then dropped the Galactic
plane and declinations $<-20^\circ$ in order to increase the cadence for the extragalactic
fields. As the system approached its current state, we resumed
observing the Northern Galactic plane. Hence the time spanned by 
the data is relatively uniform above declination $-20^\circ$, but 
there are fewer epochs along the plane.  From CTIO, we initially 
avoided the Southern Galactic plane, but then began observing it
somewhat after we resumed observing the Northern Galactic plane.
The Magellanic Clouds, M~31 and M~33 are 
observed at a higher cadence.  Most of the other smaller scale
structures are minor artifacts from dealing with some shifts in
field definitions as cameras were added.  Under the present
plan, the McDonald Observatory station will have cameras
{\tt bi}-{\tt bl}, the SAAO station will have cameras
{\tt bm}-{\tt bp}, and the second CTIO station will have
cameras {\tt bq}-{\tt bt}.  Images from the new stations will be 
incorporated into this tool after an initial period for testing.

The present tool, {\tt https://asas-sn.osu.edu/},  takes as input a coordinate (RA/Dec) and a look-back
time (days) and then provides a light curve for that position both as a graph and as a 
down-loadable table by doing aperture photometry
on the individual ASAS-SN images.  The look-back time is provided
to allow the rapid extraction of the recent behavior of
a target.  Images taken in particularly poor conditions, that
were out of focus ($FWHM > 2.5$~pixels), had poor astrometry, 
or where the source is
within $0.2$~deg of the detector edge are automatically rejected.
The latter restriction opens no gaps in the sky coverage because
of the $0.5$~deg field overlaps.  Photometry is simply extracted
at the requested position, as there is no need for any re-centering
step for the images with good astrometry.  This also means that it
is possible to obtain light curves for faint sources close to bright
sources, making allowance for the effects of crowding (see below).

For the selected images, the photometry is done using the {\tt IRAF apphot} 
package and calibrated using the AAVSO Photometric All-Sky Survey 
(APASS, \citealt{Henden2012}).  The signal is taken from a 2 pixel
radius aperture (i.e., about $2$ FHWM in diameter) and the background
is estimated in a 7 to 10-pixel radius annulus. The 
background pixels are clipped at $2\sigma$. The aperture
photometry is done both for the target position and 100
nearby $11.5 < V < 14$~mag APASS stars with photometric
uncertainties less than $0.075$~mag.  The APASS magnitude
range is chosen to avoid saturation and minimize crowding.
The APASS stars are also required to have no other APASS star
within $56\farcs25$ (about $3$ ASAS-SN FWHM) in separation
and $5$~mag in flux.

The calibration is set by the median difference between the
APASS and instrumental aperture magnitudes for each image after 
iteratively clipping the APASS calibration stars at a
threshold where we would expect to lose one or fewer stars
given Gaussian errors.  We assume that any dispersion larger
than $0.15$~mag is dominated by outliers and use this as a
maximum for the estimated dispersion.  After clipping the
calibration sample, the weighted mean is used as the final
calibration.  The returned light curve gives the HJD and
UT dates of the observation, the camera name ({\tt b?}), the
FWHM (for monitoring any residual focus issues), the estimated
$5\sigma$ detection limit for the location, the aperture magnitude
and its uncertainty and the flux estimate and its uncertainty 
(in mJy). The aperture magnitude becomes the $5\sigma$ detection
limit once it is below the detection limit.  However, the fluxes
are always the real flux measurements and their uncertainties
and so can be easily combined across images even when the 
magnitude is only reported as a limit.

We illustrate the utility of ASAS-SN light curves with four examples
of non-transient sources.  In each case, these
are the light curves for the individual exposures, and the
multiple exposures for each epoch can be combined to reduce the 
uncertainties or improve the upper limits.  We simply show
the data as returned by the query.
Figure~\ref{fig:rcorbor} shows the ASAS-SN light curve of a
classic example of a high amplitude variable star, R Coronae
Borealis (R CrB).
The light curve spans 7.3~mag and is
well-defined even at its brightest points, where it is well
above ASAS-SN's nominal saturation magnitude (see below).
Figure~\ref{fig:mdwarf} shows a much more subtle example,
the phased light curve of the eclipsing binary M dwarfs 
KELT~J041621$-$620046 (\citealt{Lubin2017}).  This $1.11$ day binary 
has a mean magnitude of $V\simeq 13.9$ and $0.3$~mag eclipses 
that are easily seen in the ASAS-SN data.  The outliers 
in Figure~\ref{fig:mdwarf} can be eliminated by rejecting
data with unusually large magnitude uncertainties.  As a more distant
example, Figure~\ref{fig:tzo} shows the ASAS-SN light curve of HV~2112,
a star in the Small Magellanic Cloud that \cite{Levesque2014} 
propose is a Thorne-Zytkow object (a red supergiant with a neutron star at its core).
Finally, as an extragalactic example, Figure~\ref{fig:ngc5548}
shows the ASAS-SN light curve of the classic reverberation mapping
target (most recently by \citealt{DeRosa2015}), NGC~5548.  The
variability of this AGN is well determined despite doing the
photometry at the center of a resolved galaxy.

The two main issues the user should be aware of are crowding and
saturation.  The 
ASAS-SN light curves are fairly robust against both problems,
provided they are neither pushed to extremes nor over-interpreted.  
To help visually evaluate both issues, the light curve server page 
returns an Aladin Sky Atlas (\citealt{Bonnarel2000}, \citealt{Boch2014})
image of the region surrounding the target. This also provides
a visual confirmation of the selected target.

An example of crowding is the massive, eclipsing Wolf-Rayet
binary WR20a (\citealt{Bonanos2004}).  Figure~\ref{fig:crowd1} shows the 
ASAS-SN and DSS images of WR20a, where the circle around the source 
is 10\farcs0 in radius.  At the resolution of ASAS-SN, the local crowding 
will tend to change the light curve by some constant in flux, distorting
its shape.  Figure~\ref{fig:crowd2} shows the phased ASAS-SN light 
curve of WR20a, and we find that the eclipse depths are
slightly underestimated (about $0.3$ mag instead of $0.4$ mag)
because of the crowding.  Since the ASAS-SN image FWHM are set by 
the camera optics rather than atmospheric seeing, there are essentially
no effects equivalent to the light curve distortions produced by combining
crowding with variable seeing.  Despite the distortion 
of the light curve, there is no difficulty recognizing that the source is an
eclipsing binary and determining the period.  In fact, Figure~\ref{fig:crowd2}
used the primary eclipse time of $2453124.569$ from \cite{Bonanos2004}
but a revised period of 3.684599~days because the original $3.686\pm0.01$~day
estimate clearly led to small phasing errors.

ASAS-SN saturates at 10-11~mag, where the exact limit depends on the
camera and the image position (vignetting).  However, ASAS-SN uses a procedure
inherited from the original ASAS survey (\citealt{Pojmanski2002}) that
moves flux out of the bleed trails of bright stars and back
into the central image. After basic processing, frames are scanned for
saturated pixels.  Once one is found, all connected saturated pixels
and their associated edge effects are identified.  The flux
of the saturated pixels is then added back as a Gaussian
around the centroid position.  The regions from which the flux was
removed are filled in by linear interpolation of the adjacent, 
unsaturated pixels.  

\begin{figure*}[t]
\centering
\includegraphics[height=2.8in]{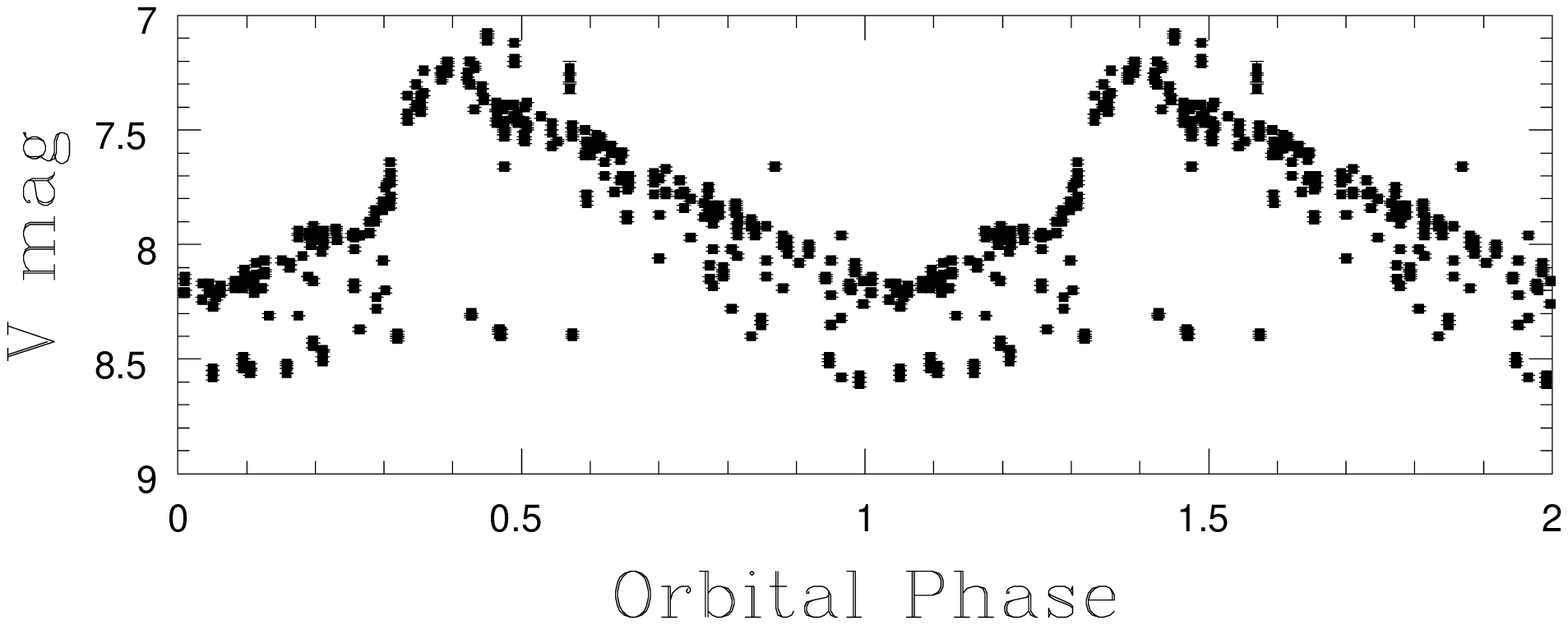}
\caption{ Phased ASAS-SN light curve of the bright, saturated $P=18.9$~day
  Cepheid VY~Car.  There are 20 points that lie below $9$~mag.
  While there are clear outliers, the light curve is surprisingly good
  given that the star is $\sim10$ times brighter than the saturation limit.
  With some straightforward editing of the un-phased data, a much cleaner
  light curve can be constructed.
  }
\label{fig:vycar}
\end{figure*}

To the extent that the charge bleeding process is conservative and the 
saturated star is relatively isolated, this leads to a substantial 
improvement in the photometry of saturated stars.  We illustrate
this in Figure~\ref{fig:vycar} with the light curve of the 
$P=18.9$ day Cepheid VY~Car.  There are clearly 
outliers and it would be unwise to use the photometry for
determining the distance scale, but
there is no difficulty recognizing the classic, long-period
Cepheid light curve and determining its period and phase. There 
are also some clear patterns in the un-phased light curves that 
would allow the production of a clean light curve after
some manual editing.  This bright
star correction procedure will not, however, salvage the light curves 
of the stars that lay underneath the bleed trails.

\section{Discussion}

While {\tt https://asas-sn.osu.edu/} as V1.0 of ASAS-SN light curve 
servers is slow because the 
light curves must be computed upon request, it does provide the 
first astronomical resource able to provide an up to date 
light curve for any 
point on the celestial sphere upon demand.  If the requested
position is currently visible, the most recent epochs will 
generally be less than a week old even with weather interruptions. 
This service is not meant for obtaining large numbers of light 
curves, with requests being self-limiting by the allocated 
computational resources.  Requests for recent behavior will
be much faster than requests for full light curves.  For one random
example, it took 10, 33, 90 and 130~sec to obtain a light
curve for the last 20 days, 100~days, 1000~days and the
full data set, respectively.  Astronomers interested in obtaining
large numbers of light curves should contact the ASAS-SN team.  

As outlined in the introduction, the next planned step is to provide
light curves for known variables and variables newly discovered by
ASAS-SN.  This will be significantly faster, since the light curves
will be pre-computed and stored in a database, but restricted to
a particular catalog of objects.  This is a well-defined
process and should be completed in 2018.  The full scope of the third
phase, supplying light curves of sources in general, has yet to
be fully determined.

\section*{Acknowledgments}

We thank the Las Cumbres Observatory and its staff for its continuing
support of the ASAS-SN project. We thank M. Hardesty of the OSU ASC
technology group.

ASAS-SN is supported by the Gordon and Betty Moore Foundation through grant 
GBMF5490 to the Ohio State University and NSF grant AST-1515927. 
Development of ASAS-SN has been supported by NSF grant AST-0908816, 
the Mt. Cuba Astronomical Foundation, 
the Center for Cosmology and AstroParticle Physics at the Ohio State University, 
the Chinese Academy of Sciences 
South America Center for Astronomy (CASSACA), the Villum
Foundation and George Skestos.

KZS, CSK and TAT are supported by NSF grants AST-1515927 and AST-1515876.  
BJS is supported by NASA through Hubble Fellowship grant HST-HF-51348.001 awarded 
by the Space Telescope Science Institute, which is operated by the Association 
of Universities for Research in Astronomy, Inc., for NASA, under contract NAS 5-26555. 
TW-SH is supported by the DOE Computational Science Graduate Fellowship, 
grant number DE-FG02-97ER25308. 

Support for JLP is in part provided by FONDECYT through the grant 1151445 
and by the Ministry of Economy, Development, and Tourism's Millennium 
Science Initiative through grant IC120009, awarded to The Millennium 
Institute of Astrophysics, 
SD is supported by Project 11573003 supported by NSFC. 

This research was made possible through the use of the AAVSO Photometric All-Sky 
Survey (APASS), funded by the Robert Martin Ayers Sciences Fund.
This research has made use of data provided by Astrometry.net 
(\citealt{Lang2010}).  This research has made use of the ``Aladin sky atlas" 
developed at CDS, Strasbourg 
Observatory, France (\citealt{Bonnarel2000}, \citealt{Boch2014}).


\begin{thebibliography}{}

\bibitem[Aartsen et al.(2017)]{Aartsen2017} Aartsen, M.~G., Ackermann, M., Adams, J., et al.\ 2017, arXiv:1702.06131 

\bibitem[Abeysekara et al.(2015)]{Abseysekara2015} Abeysekara, A.~U., Archambault, S., Archer, A., et al.\ 2015, \apjl, 815, L22 



\bibitem[Boch \& Fernique(2014)]{Boch2014} Boch, T., \& Fernique, P.\ 2014, Astronomical Data Analysis Software and Systems XXIII, 485, 277 

\bibitem[Bonanos et al.(2004)]{Bonanos2004} Bonanos, A.~Z., Stanek, K.~Z., Udalski, A., et al.\ 2004, \apjl, 611, L33 


\bibitem[Bonnarel et al.(2000)]{Bonnarel2000} Bonnarel, F., Fernique, P., Bienaym{\'e}, O., et al.\ 2000, \aaps, 143, 33 


\bibitem[Brown et al.(2013)]{Brown2013} Brown, T.~M., Baliber, N., Bianco, F.~B., et al.\ 2013, \pasp, 125, 1031 

\bibitem[Brown et al.(2016)]{Brown2016} Brown, J.~S., Shappee, B.~J., Holoien, T.~W.-S., et al.\ 2016, \mnras, 462, 3993 

\bibitem[Brown et al.(2017a)]{Brown2017a} Brown, J.~S., Holoien, T.~W.-S., Auchettl, K., et al.\ 2017a, \mnras, 466, 4904 

\bibitem[Brown et al.(2017b)]{Brown2017b} Brown, J.~S., Kochanek, C.~S., Holoien, T.~W.-S., et al.\ 2017b, arXiv:1704.02321 

\bibitem[Davis et al.(2015)]{Davis2015} Davis, A.~B., Shappee, B.~J., Archer Shappee, B., \& ASAS-SN 2015, American Astronomical Society Meeting Abstracts, 225, 344.02 

\bibitem[De Rosa et al.(2015)]{DeRosa2015} De Rosa, G., Peterson, B.~M., Ely, J., et al.\ 2015, \apj, 806, 128 


\bibitem[Godoy-Rivera et al.(2017)]{Godoy2017} Godoy-Rivera, D., Stanek, K.~Z., Kochanek, C.~S., et al.\ 2017, \mnras, 466, 1428 
\bibitem[Dong et al.(2016)]{Dong2016} Dong, S., Shappee, B.~J., Prieto, J.~L., et al.\ 2016, Science, 351, 257 

\bibitem[Drake et al.(2009)]{Drake2009} Drake, A.~J., Djorgovski, S.~G., Mahabal, A., et al.\ 2009, \apj, 696, 870 
\bibitem[Gully-Santiago et al.(2017)]{Gully2017} Gully-Santiago, M.~A., Herczeg, G.~J., Czekala, I., et al.\ 2017, \apj, 836, 200 


\bibitem[Henden et al.(2012)]{Henden2012} Henden, A.~A., Levine, S.~E., Terrell, D., Smith, T.~C., \& Welch, D.\ 2012, Journal of the American Association of Variable Star Observers (JAAVSO), 40, 430 

\bibitem[Herczeg et al.(2016)]{Herczeg2016} Herczeg, G.~J., Dong, S., Shappee, B.~J., et al.\ 2016, \apj, 831, 133 

\bibitem[Holoien et al.(2014a)]{Holoien2014a} Holoien, T.~W.-S., Prieto, J.~L., Stanek, K.~Z., et al.\ 2014a, \apjl, 785, L35 
\bibitem[Holoien et al.(2014b)]{Holoien2014b} Holoien, T.~W.-S., Prieto, J.~L., Bersier, D., et al.\ 2014b, \mnras, 445, 3263 
\bibitem[Holoien et al.(2016a)]{Holoien2016a} Holoien, T.~W.-S., Kochanek, C.~S., Prieto, J.~L., et al.\ 2016a, \mnras, 455, 2918 

\bibitem[Holoien et al.(2016b)]{Holoien2016b} Holoien, T.~W.-S., Kochanek, C.~S., Prieto, J.~L., et al.\ 2016b, \mnras, 463, 3813 

\bibitem[Holoien et al.(2017a)]{Holoien2017a} Holoien, T.~W.-S., Stanek, K.~Z., Kochanek, C.~S., et al.\ 2017a, \mnras, 464, 2672 

\bibitem[Holoien et al.(2017b)]{Holoien2017b} Holoien, T.~W.-S., Brown, J.~S., Stanek, K.~Z., et al.\ 2017b, \mnras, 467, 1098 

\bibitem[Holoien et al.(2017c)]{Holoien2017c} Holoien, T.~W.-S., Brown, J.~S., Stanek, K.~Z., et al.\ 2017c, arXiv:1704.02320 

\bibitem[Ivezi{\'c} et al.(2007)]{Ivezic2007} Ivezi{\'c}, {\v Z}., Smith, J.~A., Miknaitis, G., et al.\ 2007, \aj, 134, 973 

\bibitem[Kato et al.(2016)]{Kato2016} Kato, T., Hambsch, F.-J., Monard, B., et al.\ 2016, \pasj, 68, 65 

\bibitem[Lang et al.(2010)]{Lang2010} Lang, D., Hogg, D.~W., Mierle, K., Blanton, M., \& Roweis, S.\ 2010, \aj, 139, 1782 

\bibitem[Law et al.(2009)]{Law2009} Law, N.~M., Kulkarni, S.~R., Dekany, R.~G., et al.\ 2009, \pasp, 121, 1395 

\bibitem[Levesque et al.(2014)]{Levesque2014} Levesque, E.~M., Massey, P., {\.Z}ytkow, A.~N., \& Morrell, N.\ 2014, \mnras, 443, L94 

\bibitem[Littlefield et al.(2016)]{Littlefield2016} Littlefield, C., Garnavich, P., Kennedy, M.~R., et al.\ 2016, \apj, 833, 93 

\bibitem[Lubin et al.(2017)]{Lubin2017} Lubin, J.~B., Rodriguez, J.~E., Zhou, G., et al.\ 2017, arXiv:1706.02401 


\bibitem[Osborn et al.(2017)]{Osborn2017} Osborn, H.~P., Rodriguez, J.~E., Kenworthy, M.~A., et al.\ 2017, arXiv:1705.10346 

\bibitem[Paczy{\'n}ski(2000)]{Paczynski2000} Paczy{\'n}ski, B.\ 2000, \pasp, 112, 1281 


\bibitem[Pojmanski(2002)]{Pojmanski2002} Pojmanski, G.\ 2002, Acta A., 52, 397 

\bibitem[Rodriguez et al.(2016)]{Rodriguez2016} Rodriguez, J.~E., Stassun, K.~G., Cargile, P., et al.\ 2016, \apj, 831, 74 

\bibitem[Rodriguez et al.(2017)]{Rodriguez2017} Rodriguez, J.~E., Zhou, G., Cargile, P.~A., et al.\ 2017, \apj, 836, 209 

\bibitem[Schmidt et al.(2014)]{Schmidt2014} Schmidt, S.~J., Prieto, J.~L., Stanek, K.~Z., et al.\ 2014, \apjl, 781, L24 

\bibitem[Schmidt et al.(2016)]{Schmidt2016} Schmidt, S.~J., Shappee, B.~J., Gagn{\'e}, J., et al.\ 2016, \apjl, 828, L22 

\bibitem[Shappee et al.(2014)]{Shappee2014} Shappee, B.~J., Prieto, J.~L., Grupe, D., et al.\ 2014, \apj, 788, 48 

\bibitem[Stanek et al.(2016)]{Stanek2016} Stanek, K.~Z., Kochanek, C.~S., Brown, J.~S., et al.\ 2016, The Astronomer's Telegram, 9669

\bibitem[Udalski et al.(2008)]{Udalski2008} Udalski, A., Szymanski, M.~K., Soszynski, I., \& Poleski, R.\ 2008, Acta A., 58, 69 

\bibitem[Wo{\'z}niak et al.(2004)]{Wozniak2004} Wo{\'z}niak, P.~R., Vestrand, W.~T., Akerlof, C.~W., et al.\ 2004, \aj, 127, 2436 

\end{thebibliography}
\end{document}